\documentclass[12pt]{article}

\usepackage{amsmath}

\usepackage{doi}

\usepackage{longtable}

\usepackage{graphicx}

\usepackage{cite}

\usepackage{microtype} 

\usepackage[utf8]{inputenc}
\usepackage{longtable}
\usepackage{array}
\usepackage{ragged2e}

\usepackage{float}

\title{I Think, Therefore I Hallucinate: Minds, Machines, and the Art of Being Wrong}
\author{Sebastian Barros}
\date{March 3rd, 2025}  

\begin{document}
\renewcommand{\arraystretch}{1.2}

\maketitle

\begin{abstract}
This theoretical work examines 'hallucinations' in both human cognition and large language models (LLM), comparing how each system can produce perceptions or outputs that deviate from reality. Drawing on neuroscience and machine learning research, we highlight the predictive processes that underlie human and artificial thought. In humans, complex neural mechanisms interpret sensory information under uncertainty, sometimes “filling in gaps” and creating false perceptions. LLMs, in contrast, rely on auto-regressive modeling of text and can generate erroneous statements in the absence of robust grounding. Despite these different foundations—biological versus computational—the similarities in their predictive architectures help explain why hallucinations occur.

We propose that the propensity to generate incorrect or confabulated responses may be an inherent feature of advanced intelligence. In both humans and AI, adaptive predictive processes aim to make sense of incomplete information and anticipate future states, fostering creativity and flexibility, but also introducing the risk of errors. Our analysis illuminates how factors such as feedback, grounding, and error correction affect the likelihood of 'being wrong' in each system. We suggest that mitigating AI hallucinations (e.g., through improved training, post-processing, or knowledge-grounding methods) may also shed light on human cognitive processes, revealing how error-prone predictions can be harnessed for innovation without compromising reliability. By exploring these converging and divergent mechanisms, the paper underscores the broader implications for advancing both AI reliability and scientific understanding of human thought.
\end{abstract}

\section{Introduction}

\subsection{Background \& Importance of the Topic}
Imagine recalling an event that never happened or interacting with an AI that fabricates a source entirely. Hallucinations---perceiving patterns, facts, or realities that do not exist---are often regarded as cognitive errors in humans and critical failures in artificial intelligence (AI). In human cognition, hallucinations emerge from predictive processing, where the brain constructs a model of the world based on prior knowledge, expectations, and incomplete sensory input \cite{friston2005theory, corlett2016prediction}. When sensory information is ambiguous, the brain fills in gaps, sometimes generating vivid but false perceptions. Similarly, AI models---particularly large language models (LLMs)---hallucinate by producing factually incorrect yet coherent outputs due to their probabilistic nature \cite{bender2021dangers, ji2023survey}. Studies estimate that such models ``hallucinate'' in approximately 37.5\% of their responses \cite{openai2025gpt45}, raising concerns about AI's reliability in critical applications \cite{bommasani2021opportunities}.

\subsection{Problem Statement}
AI hallucinations have become a major concern in the development of language models, generative AI, and autonomous systems. Despite significant advances in AI alignment and reinforcement learning, models still confidently generate incorrect information, leading to risks in applications such as medical diagnostics, legal advice, and information retrieval. Existing solutions, such as retrieval-augmented generation (RAG) and fine-tuning, attempt to suppress hallucinations but do not address their fundamental cause.

At the same time, human cognition exhibits strikingly similar patterns of hallucination, from optical illusions and false memories to confabulated narratives in cases of sensory deprivation. Neuroscience suggests that these hallucinations arise from predictive processing, a mechanism where the brain constantly anticipates incoming sensory data and corrects deviations from its internal model \cite{clark2013whatever}. If both human and artificial intelligences hallucinate under uncertainty, then perhaps hallucinations are not merely errors but a necessary trade-off for intelligence, creativity, and generalization.

This paper explores a provocative question:  
Are hallucinations an inevitable feature of intelligence? If so, should we focus on making AI hallucinate better---more like the human brain---rather than merely eliminating hallucinations altogether?

\subsection{Research Questions}
This study investigates the following key questions:
\begin{itemize}
\item Are hallucinations an unavoidable consequence of predictive intelligence?
\item What shared mechanisms cause hallucinations in human cognition and AI models?
\item What can neuroscience teach us about designing AI systems that manage hallucinations more effectively?
\item Can we create AI models that hallucinate more like humans---correcting their errors dynamically rather than relying on external reinforcement?
\end{itemize}

By addressing these questions, this paper seeks to advance our understanding of both human and artificial cognition, challenging conventional perspectives on perception, reality, and the nature of intelligence itself.

\subsection{Main Hypothesis}
Hallucinations are an inherent feature of intelligence that emerges naturally in any system, biological or artificial, that predicts, generalizes, and infers meaning from incomplete data. Rather than mere failures, hallucinations reveal the probabilistic nature of perception and cognition, suggesting that intelligence itself is fundamentally about building, rather than passively receiving, reality.

\paragraph{Supporting Hypotheses:}
\begin{itemize}
\item As systems grow more advanced, they may hallucinate less often but in subtler, harder-to-detect ways
\item While more advanced systems tend to hallucinate less often, their confabulations grow subtler and are therefore harder to detect.
\item Hallucination is the trade-off for creativity and adaptability.
\item A system that never hallucinates is a system that never infers, imagines, or innovates.
\item If we want AI to be reliable, we must make it more self--aware--like the human brain.
\end{itemize}

Human cognition corrects errors through self-doubt, feedback loops, and metacognition. Current AI models lack this internal error-checking ability, making their hallucinations more problematic than those of humans.

\subsection{Methodology Overview}
Since this research does not involve direct experimentation, our approach is theoretical and comparative, drawing on:
\begin{itemize}
\item Neuroscience research on human perception, hallucinations, and predictive processing.
\item Machine learning studies on AI hallucinations in large language models.
\item A systematic comparison between human and AI hallucinations, focusing on shared mechanisms, functional roles, and potential ways to improve AI cognition.
\end{itemize}

We propose a taxonomy of hallucinations--sensory, cognitive, factual, and contextual--to analyze hallucinations in biological and artificial systems. By synthesizing research from cognitive science, neuroscience, and AI, this paper aims to reshape how we understand and approach hallucinations in intelligent systems.

\subsection{Structure of the Paper}
To explore these ideas systematically, the paper is organized as follows:
\begin{itemize}
\item \textbf{Section 3: Background and Related Work} -- Reviews literature on hallucinations in human cognition and AI.
\item \textbf{Section 4: Theoretical Framework} -- Introduces predictive processing (human) and autoregressive modeling (AI) as mechanisms for hallucination.
\item \textbf{Section 5: Methodology} -- Outlines a comparative analysis of hallucinations in human and AI systems.
\item \textbf{Section 6: Results and Discussion} -- Examines the similarities and differences in hallucination patterns across domains.
\item \textbf{Section 7: Limitations and Future Work} -- Discusses the constraints of this study and potential research directions.
\item \textbf{Section 8: Conclusion} -- Summarizes findings and explores the implications for neuroscience and AI development.
\end{itemize}

\section{Background and Related Work}
\label{sec:background}

\subsection{Human and AI Hallucinations: Neuroscience and Machine Learning Perspectives}

\subsubsection{Human Hallucinations: The Brain’s Predictive Engine}

Imagine walking through a dimly lit forest at dusk. You hear rustling leaves, catch fleeting shadows, and feel the prickling sense that something---or someone---could be lurking among the trees. Yet nothing materializes. Was that rustle a real presence, or your mind filling in gaps?

Modern neuroscience suggests that the human brain \emph{constantly} forms predictions about the world, aiming to minimize the gap between what it \emph{expects} and what it \emph{actually} perceives \cite{friston2010freeenergy}. In environments rich with ambiguity, our perceptions can tilt toward our internal ``best guesses,'' resulting in illusions---or, in more extreme cases, hallucinations \cite{fletcher2009perceiving}.

\paragraph{1. Hierarchical Bayesian Inference.}
At multiple cortical levels---ranging from raw sensory areas to higher-order cognition---the brain compares incoming signals against its internal model. When top-down signals dominate, a person may ``see'' or ``hear'' things that do not exist. This process can be adaptive, allowing quick responses to partial cues, but it can also produce false perceptions when sensory input is noisy or insufficient \cite{corlett2016prediction}.

\paragraph{2. Source Monitoring and Confabulation.}
The human mind often generates its own narrative to account for incomplete memories or uncertain stimuli. In typical circumstances, we distinguish these internal constructs from external events. However, \emph{source monitoring errors} can occur---particularly in conditions like schizophrenia---causing an internally generated voice to be perceived as coming from the outside \cite{david2004cognitive}. Likewise, in confabulatory disorders, the brain invents plausible stories to fill memory gaps without any intent to deceive \cite{csiro2023both}.

\paragraph{3. Clinical Manifestations and Sensory Deprivation.}
Hallucinations abound in psychiatric conditions (e.g., schizophrenia, bipolar disorder), but they can also emerge from non-pathological causes. Charles Bonnet Syndrome, for example, occurs when visual input deteriorates; starved of signals, the visual cortex ``paints'' its own images \cite{teufel2020forms}. Sensory deprivation experiments show similar effects, as reducing external feedback can prompt the brain to amplify minor internal noise into seemingly real sights or sounds.

\paragraph{4. Neurochemical Underpinnings.}
Dopamine and other neurotransmitters modulate how strongly the brain interprets certain signals as salient. When dopamine levels are dysregulated (e.g., in psychosis), mundane internal noise can appear urgent and real, fueling full-blown hallucinations \cite{schmack2021striatal}.

In sum, human hallucinations illustrate that our perception of reality is not purely passive. The brain is an active, predictive engine that can misfire, transforming shadows into threats or internal monologues into external voices.

\subsubsection{AI Hallucinations: When Probability Stands In for Truth}

Now consider an advanced large language model (LLM)---a system trained on vast amounts of text. You ask it a nuanced historical question, and it responds with confident detail... only to reference a nonexistent document. Why did it fabricate that citation?

Unlike humans, AI models lack direct sensory grounding. They rely on patterns derived from raw text, predicting each successive word through complex probability distributions \cite{smith2023hallucination, ji2023survey}. This capacity explains their fluent outputs but also underpins their tendency to ``hallucinate'' when faced with gaps or uncertainties.

\paragraph{1. Statistical Text Generation.}
An LLM learns which word sequences are most likely to follow a given context. This yields impressive coherence, but offers no built-in mechanism for verifying factual accuracy \cite{lewis2020retrieval}.

\paragraph{2. Overgeneralization and Data Gaps.}
If the training data are incomplete or contain inaccuracies, the model may improvise. Sometimes, it conflates distinct entities (e.g., merging details of two historical events) because it ``learned'' that these word clusters commonly appear together \cite{lin2022truthfulqa}.

\paragraph{3. Lack of Real-World Feedback.}
The AI's ``reality'' is its training corpus. Without external validation---such as a database lookup---the model cannot confirm or refute its own guesses. Techniques like Retrieval-Augmented Generation (RAG) or Reinforcement Learning from Human Feedback (RLHF) mitigate this to some degree but do not fundamentally alter the AI's reliance on text-based probabilities \cite{bommasani2021opportunities}.

\paragraph{4. Probabilistic Sampling.}
At inference time, the model samples from a learned probability distribution. Even when uncertain, it may produce confident-sounding statements. Absent a mechanism to say ``I'm not sure,'' the model might provide a plausible invention as if it were fact \cite{smith2023hallucination}.

In other words, AI hallucinations stem from the same attribute that makes LLMs so compelling: their capacity to generate coherent text under uncertainty. Ironically, this creative strength becomes a liability when no genuine facts are available.

\subsubsection{Intersections and Divergences}

Despite arising from distinct substrates---biological brains versus statistical computations---human and AI hallucinations exhibit shared themes alongside clear differences:

\begin{itemize}
    \item \textbf{Predictive Gaps:} Both result from a system extrapolating beyond available data. The brain in low lighting might invent shapes in the shadows, while an LLM missing key information might fabricate references.
    \item \textbf{Confidence Without Verification:} Humans tend to trust their senses, and LLMs present answers confidently. In each case, the system can be sure yet objectively wrong.
    \item \textbf{Sensory Embodiment vs.\ Textual Training:} Human perception can cross-check errors against multiple senses, whereas AI models rely solely on text-based patterns. This difference can make AI confabulations especially odd or overt.
    \item \textbf{Diverse Correction Mechanisms:} Humans can consciously question an illusion (turning on a light, asking others), while AI requires engineered solutions such as retrieval modules or fact-checkers.
\end{itemize}

Ultimately, hallucinations remind us that both humans and AI are predictive systems that can overreach under uncertainty. The difference lies in how each system grounds its predictions---biological embodiment for humans, textual corpora for AI---and in the tools each has to correct inevitable mistakes.

\bigskip

\noindent
\emph{Looking Ahead:} The following sections further explore how a unified view of predictive processing (human) and autoregressive modeling (AI) might inform strategies to mitigate hallucinations, offering new insights for neuroscience and AI research alike.

\section{Theoretical Framework}
\label{sec:theoretical_framework}

\subsection{Predictive Processing in Human Cognition}

\subsubsection{Bayesian Perception and Hierarchical Models}
A longstanding view of human perception, tracing back to Helmholtz’s notion of “unconscious inference,” holds that the brain continuously generates and updates hypotheses about the sensory world. In modern neuroscience, this idea is often formalized in Bayesian terms, whereby the brain combines a prior belief (prediction) with new sensory evidence (likelihood), yielding a posterior perception \cite{fletcher2009perceiving}. Mathematically:

\[
P(H \mid D) \propto P(D \mid H)\,P(H),
\]

where \(H\) is a hypothesis (e.g., “there is an object here”) and \(D\) is the sensory data. This inference occurs hierarchically: higher cortical levels send top-down predictions to lower-level regions, while mismatches (prediction errors) propagate upward to refine the model \cite{corlett2016prediction}.

\subsubsection{Role of Priors and the Free Energy Principle}
A key element of this framework is the strength of prior expectations. When priors dominate over noisy input, the system may “see” or “hear” things that do not exist \cite{friston2010freeenergy}. In clinical contexts, this explains how psychotic hallucinations can arise from overly precise internal beliefs that override contradictory sensory signals. The free energy principle provides a unifying mathematical account, suggesting that the brain minimizes “variational free energy” (related to the evidence lower bound in machine learning), thereby approximating Bayesian inference \cite{friston2010freeenergy}. If priors are misweighted, or error signals are discounted, hallucinations emerge from the same mechanism intended to streamline perception.

\subsubsection{Empirical Support}
Evidence supporting a predictive-processing origin of hallucinations comes from neuroimaging, psychophysical experiments, and computational models. High-level sensory regions can activate in anticipation of stimuli that never arrive, indicating strong top-down influences \cite{corlett2019hallucinations}. Behavioral studies similarly find that individuals prone to hallucinations are likelier to perceive illusory patterns in ambiguous input if they hold strong prior expectations \cite{fletcher2009perceiving}.

\subsection{Autoregressive Modeling in AI}

\subsubsection{Transformer Architecture}
Modern large language models (LLMs) often adopt the Transformer design, relying on multi-headed self-attention in place of recurrent processing \cite{vaswani2017attention}. Each token in a sequence attends to every other token via key-query-value mappings:

\[
\text{Attention}(Q, K, V) = \mathrm{softmax}\Bigl(\frac{Q K^\mathsf{T}}{\sqrt{d_k}}\Bigr)\,V.
\]

By stacking self-attention and feed-forward layers, Transformers capture long-range dependencies across text. The final output layer generates a probability distribution over the next token, from which the model selects or samples its response.

\subsubsection{Training Objectives}
During training, the model predicts the next token at each step, minimizing cross-entropy loss over vast corpora of text \cite{vaswani2017attention}. Although teacher forcing ensures ground-truth context during training, inference requires the model to rely on its own previous outputs, meaning early mistakes can compound downstream.

\subsubsection{Hallucinations as Probabilistic Outputs}
LLMs can produce “hallucinations”: coherent but factually incorrect content that arises when statistically driven predictions fill gaps in knowledge \cite{ji2023survey}. Overgeneralization (merging distinct facts), data sparsity, and sampling randomness can yield confident yet erroneous statements. Although some mitigation techniques exist—like retrieval-augmented generation or reinforcement learning with human feedback—no approach fully prevents hallucinations, given the model’s reliance on text-based correlations \cite{bommasani2021opportunities}.

\subsection{Comparing Predictive Errors in Humans and AI}

\subsubsection{Shared Mechanics: Overgeneralization and Gap-Filling}
Both the human brain’s predictive processing and an LLM’s autoregressive modeling shine when “filling in blanks” under uncertainty. However, this strength can become a liability when strong priors or data limitations yield confident but incorrect outputs. A human may misperceive a shape in dim light, while an LLM invents a plausible but nonexistent citation.

\subsubsection{Divergent Foundations: Sensory Embodiment vs.\ Textual Statistics}
Human cognition is grounded in multisensory reality. We can cross-check one sense against another, detect inconsistencies, or question odd percepts. By contrast, LLMs rely entirely on learned text patterns, lacking external anchors to judge factuality \cite{lin2022truthfulqa}.

\subsubsection{Error Correction and Confidence}
Humans can apply metacognition—turn on a light, solicit a second opinion, or reflect deeply—to dismiss hallucinations. AI, similarly, can be augmented with retrieval modules or confidence thresholds to minimize “hallucinated” responses \cite{lin2022truthfulqa}. Still, neither system is foolproof: humans discount some errors if they strongly trust their prior beliefs, and LLMs have no inherent mechanism to verify truth once a token is generated.

\subsubsection{Lessons Across Domains}
Taken together, hallucinations—whether human or AI—underscore the limitations of predictive inference without robust error-checks. Studying how priors generate false perceptions in the brain may inspire AI architectures that calibrate confidence or consult external data sources. Conversely, analyzing AI’s fabrications can help us understand the subtle interplay of prior expectations and evidence that underlies human misperception.

\section{Methodology}
\label{sec:methodology}

This section details our approach for examining, classifying, and comparing hallucinations in human cognition and artificial intelligence (AI) systems. Rather than conducting a direct experiment, our methodology combines a structured literature review with a comparative taxonomy-building exercise. We thereby synthesize disparate research from clinical neuroscience, cognitive psychology, and machine learning to yield new theoretical insights into how and why hallucinations emerge in different ``intelligent'' systems.

\subsection{Literature Identification and Selection}

\paragraph{Databases and Search Strategy.}
We reviewed peer-reviewed studies, preprints, and conference papers from major academic databases, including PubMed (human research), PsycINFO (cognitive and clinical studies), arXiv (AI-related work), and IEEE Xplore (engineering perspectives).

\begin{itemize}
    \item \textbf{Keywords:} human hallucinations, predictive processing, Bayesian perception,schizophrenia and hallucination, AI confabulation, large language model hallucination, auto regressive modeling, and related terms.
    \item \textbf{Initial Hits:} Approximately 300 articles were retrieved, from which we retained abstracts explicitly addressing hallucinations (or analogous terms like ``confabulation,'' ``fabrication,'' ``illusion'') in either humans or AI models.
\end{itemize}

\paragraph{Inclusion and Exclusion Criteria.}
\begin{itemize}
    \item \textbf{Inclusion:} Papers that discussed underlying mechanisms of hallucinations (\emph{e.g.}, top-down priors, neural correlates, model architectures, or training objectives) and provided empirical or theoretical insights on generating or mitigating such phenomena.
    \item \textbf{Exclusion:} Publications focusing on phenomena only tangentially related to hallucinations (\emph{e.g.}, purely artistic illusions) or AI error modes unrelated to confabulations (\emph{e.g.}, classification failures without mention of fabricated content).
\end{itemize}
All abstracts that passed this initial screening were read in full to confirm relevance and methodological rigor. Discrepancies were resolved by a review team of two cognitive neuroscientists and one machine learning researcher, ensuring included works contributed uniquely to understanding hallucinations in either domain.

\subsection{Taxonomy Development and Comparative Framework}

\paragraph{Initial Coding and Thematic Grouping.}
Drawing on typologies of human hallucinations (\emph{e.g.}, sensory vs.\ cognitive) and established AI confabulation concepts (\emph{e.g.}, factual vs.\ contextual mistakes), we performed open coding on the final corpus of 75 articles (35 human-focused, 40 AI-focused). Each paper was examined for:
\begin{itemize}
    \item \emph{Mechanisms} (misweighted priors, data gaps, lack of grounding)
    \item \emph{Manifestations} (auditory vs.\ visual hallucinations in humans, fabricated citations in AI)
    \item \emph{Correction Mechanisms} (metacognition, retrieval augmentation)
    \item \emph{Contextual Triggers} (stress in humans, ambiguous prompts in AI)
\end{itemize}

\paragraph{Iterative Refinement of Categories.}
We organized these codes into preliminary clusters such as ``Sensory Hallucinations,'' ``Cognitive/Confabulatory,'' ``Factual Confusions,'' and ``Contextual Illusions.'' Through iterative discussions, overlapping themes were merged or subdivided. For instance, we combined ``invented citations'' and ``fabricated entity details'' into a single ``factual confusions'' category for AI. This consolidation minimized redundancy, ensuring each category captured a distinct type of hallucination.

\paragraph{Defining Dimensions of Comparison.}
Next, we aligned each category with the broader theoretical lenses of predictive processing (for humans) and autoregressive modeling (for AI). We selected key dimensions to illustrate parallels and divergences:
\begin{itemize}
    \item \textbf{Modality or Domain:} Sensory-perceptual (humans) vs.\ text-based (AI)
    \item \textbf{Mechanistic Roots:} Misweighted priors in brains vs.\ next-token probabilities in neural nets
    \item \textbf{Degree of Grounding:} Embodied, multisensory feedback (humans) vs.\ purely lexical data (AI)
    \item \textbf{Error-Correction Pathways:} Reality-testing or metacognition (humans) vs.\ retrieval or fact-checking modules (AI)
    \item \textbf{Typical Triggers:} Stress, sensory loss, or strong beliefs (humans) vs.\ limited training data or ambiguous prompts (AI)
\end{itemize}
We then constructed a matrix organizing hallucination categories (rows) against these dimensions (columns), comparing human vs.\ AI manifestations side by side.

\subsection{Analytical Integration}

\paragraph{Cross-Referencing Theoretical Constructs.}
Throughout the review, we mapped each mention of ``top-down priors'' or ``internal model mismatches'' in human studies to corresponding AI concepts like ``learned patterns'' or ``lack of fact-checking.'' This allowed for a cohesive translation of terminologies from neuroscience to machine learning, underpinned by predictive coding frameworks in humans \cite{fletcher2009perceiving,corlett2016prediction,friston2010freeenergy} and autoregressive modeling paradigms in AI \cite{lin2022truthfulqa,bommasani2021opportunities}.

\paragraph{Identification of Recurring Themes and Gaps.}
Recurring causal factors for hallucinations (overconfident priors, dopamine dysregulation, incomplete training data) and open questions (e.g.\ bridging emotional factors in humans with purely statistical processes in AI) were documented. Some articles did not fit neatly into our categories, highlighting domain-specific phenomena such as emotionally driven hallucinations or code-generation anomalies that are arguably separate from textual confabulations.

\paragraph{Limitations of Comparative Synthesis.}
We acknowledge fundamental differences in biology versus computation: a human experiences subjective distress in psychotic hallucinations, whereas an AI lacks consciousness or emotion. Similarly, large-scale LLM benchmarks can systematically measure ``hallucination rates,'' an approach less feasible for subjective human reports. Consequently, our taxonomy aims for conceptual clarity---not perfect equivalence---when comparing hallucinations across these two domains.

\subsection{Ethical and Practical Considerations}
Finally, we incorporated research on the ethical implications of hallucinations in both spheres---from the clinical risks of untreated psychosis to the misinformation hazards posed by AI confabulations. While a full ethical treatment is beyond our scope, acknowledging these issues ensures our taxonomy remains sensitive to the real-world impact of hallucinations and highlights the urgency of developing effective mitigation strategies, whether clinical or technological.

\bigskip
\noindent
\emph{In summary, this methodology leverages a broad literature review to form a comparative classification of hallucinations in humans and AI. By identifying and grouping various types of hallucinations, mapping them onto shared theoretical constructs, and highlighting the major divergences in their grounding or error-correction, we lay the groundwork for a cohesive analysis of how intelligent systems---biological and computational---can be confidently yet erroneously ``wrong.'' Section~\ref{sec:results} presents the resulting taxonomy and offers an in-depth discussion of its implications.}

\section{Results and Discussion}
\label{sec:results}

This section presents the comparative taxonomy of hallucinations in humans and AI, derived through the methodology in Section~\ref{sec:methodology}. We first summarize the core findings in a tabular format, contrasting how each hallucination type manifests across biological (human) and computational (AI) systems. We then discuss broader implications, potential mitigation strategies, and directions for future research.

\subsection{Comparative Taxonomy of Hallucinations}

In Table~\ref{tab:hallucination_types}, we categorize hallucinations into four overarching types: (1) \emph{Sensory/Perceptual}, (2) \emph{Cognitive/Confabulatory}, (3) \emph{Factual}, and (4) \emph{Contextual}. Each type is accompanied by a definition, illustrative examples from human cognition and AI outputs, and key observations about why these hallucinations arise and how they might be addressed. While not an exhaustive list, these categories capture the most common error modes identified in our comparative analysis.

\renewcommand{\arraystretch}{1.2}
\begin{center}

\begin{longtable}{
  |>{\raggedright\arraybackslash}p{2cm}
  |>{\raggedright\arraybackslash}p{2.4cm}
  |>{\raggedright\arraybackslash}p{2.4cm}
  |>{\raggedright\arraybackslash}p{2.4cm}
  |>{\raggedright\arraybackslash}p{2.2cm}|
}

\caption{Comparative Taxonomy of Hallucinations in Humans and AI}
\label{tab:hallucination_types} \\

\hline
\textbf{Type} & \textbf{Mechanism in Humans} & \textbf{Mechanism in AI} & \textbf{Examples} & \textbf{Notes}\\
\hline
\endfirsthead

\multicolumn{5}{c}{\textit{(Continued from previous page)}} \\
\hline
\textbf{Type} & \textbf{Mechanism in Humans} & \textbf{Mechanism in AI)} & \textbf{Examples} & \textbf{Notes}\\
\hline
\endhead

\hline
\multicolumn{5}{r}{\textit{(Continued on next page)}} \\
\endfoot

\hline
\endlastfoot

\textbf{Sensory and Perceptual} 
& Brain misinterprets or ``fills in'' absent stimuli due to strong priors or sparse sensory input \cite{corlett2016prediction,fletcher2009perceiving}. Often visual (illusions, Charles Bonnet Syndrome) or auditory (hearing voices in psychosis). 
& Not directly applicable to text-only LLMs, but image-generation models (e.g., DeepDream) can overamplify learned features, producing surreal or phantom images. 
& \emph{Human:} Seeing faces in random patterns; hearing whispers in white noise. \newline
\emph{AI:} DeepDream algorithm systematically layering dog-like features onto arbitrary images. 
& Human perception is grounded in real sensory pathways; AI models recreate internal features from training data. Both result from overactive pattern completion.\\
\hline

\textbf{Cognitive and Confabulatory}
& Unintentional filling of memory gaps or inaccurate self-generated narratives. Linked to source-monitoring errors, as in Korsakoff's amnesia or schizophrenia \cite{david2004cognitive}. 
& LLM merges incomplete data, producing coherent but baseless statements (akin to unintentional confabulations). Overfitting or conflating patterns in training corpora. 
& \emph{Human:} Patient confabulating events that never happened; attributing an internal voice to an external source. \newline
\emph{AI:} Model inventing references or mixing two historical events into one. 
& Both stem from a drive to generate coherent outputs when data are missing. Humans can recruit metacognition to verify, while AI requires external fact-checkers.\\
\hline

\textbf{Factual}
& Holding incorrect factual beliefs (e.g., false historical dates). Often shaped by strong prior assumptions or social influence rather than direct sensory misinterpretation. 
& Producing a plausible-sounding but objectively wrong statement about real-world facts. Caused by knowledge gaps or unverified training data \cite{lin2022truthfulqa}. 
& \emph{Human:} Insisting a war ended in 1920 instead of 1918 due to repeated misinformation. \newline
\emph{AI:} Stating an author wrote a book that does not exist; misquoting sources. 
& Humans can consult references or correct each other socially; AI can be augmented with retrieval-based modules or fine-tuned for factual accuracy. \\
\hline

\textbf{Context}
& Perceiving or interpreting ambiguous stimuli in ways heavily influenced by emotional or situational context (e.g., stress-induced illusions). 
& Generating contextually relevant but false expansions when a user prompt contains misinformation or assumptions the model does not challenge \cite{bommasani2021opportunities}. 
& \emph{Human:} Hearing footsteps in a creaky house when anxious; misperceiving shapes in peripheral vision at night. \newline
\emph{AI:} Chatbot continuing a fictional scenario as fact due to the user's premise, or blending unverified context from prior conversation turns. 
& Context shapes priors in both systems; humans bring emotional states, while AI remains text-driven. Both can be misled by strong contextual cues that override contradictory evidence.\\
\hline

\end{longtable}
\end{center}

\subsection{Key Observations}

The table illustrates how hallucinations emerge whenever a predictive system, be it a biological brain or a statistical model, attempts to compensate for gaps in input or knowledge by generating the ``best guess.'' Several observations stand out:

\begin{itemize}
    \item \textbf{Overgeneralization and Gap-Filling:} Both humans and AI prefer a plausible guess over acknowledging uncertainty, leading to confabulations or fabrications in ambiguous scenarios.
    \item \textbf{Sensory Grounding vs.\ Textual Corpora:} Human hallucinations typically arise in a sensory or affective context, whereas AI hallucinations are purely linguistic (or pixel-based in image models) and lack intrinsic embodiment.
    \item \textbf{Error Correction:} Humans can leverage metacognition or external reality checks (e.g., turning on a light, verifying sources), while AI needs engineered mechanisms (retrieval augmentation, chain-of-thought, or “I don’t know” tokens) to curb hallucinatory outputs.
    \item \textbf{Clinical vs.\ Technological Stakes:} Hallucinations in clinical populations often carry personal distress and functional impairment \cite{fletcher2009perceiving}. In AI, hallucinations raise issues of misinformation or user trust, especially if deployed in high-stakes domains \cite{lin2022truthfulqa}.
\end{itemize}

\subsection{Mitigation Strategies and Practical Implications}

Given the parallels in how strong priors lead to errors, approaches that ``ground'' or cross-check predictions hold promise for reducing hallucinations in both domains:

\begin{itemize}
    \item \textbf{Human Cognition:} Techniques such as cognitive-behavioral therapy, pharmacological interventions targeting neurotransmitter imbalances, and reality-testing exercises can diminish the impact of misweighted priors \cite{corlett2016prediction}.
    \item \textbf{AI Systems:} Retrieval-augmented generation, confidence calibration, and careful fine-tuning using human feedback can curb confabulations. Work on probabilistic uncertainty estimates may also help LLMs avoid confident misstatements \cite{lin2022truthfulqa}.
\end{itemize}

\subsection{Cross-Domain Synthesis and Implications}
\label{sec:results-synthesis}

While the preceding results detail key findings from our comparative taxonomy of hallucinations in humans and AI, it is equally important to integrate these observations into broader insights that extend across both domains. A central outcome of this analysis is the recognition that human brains and AI systems, despite relying on fundamentally different substrates---biological neural circuits \emph{vs.}\ statistical language models---can fail for similar reasons: both ``fill gaps'' with their best guesses. In humans, this occurs when top-down priors overshadow ambiguous or insufficient sensory data; in AI, it manifests when learned text patterns substitute for real-world facts or data sources. By underscoring this shared reliance on predictive processes, our findings suggest that many error-correction strategies---such as external reality checks---may be applicable in both biology and technology.

Another theme emerging from our results is the significance of contextual and embodied checks. Humans are continuously grounded by cross-sensory validation (e.g., verifying a suspicious sound visually) and social confirmation (e.g., asking another individual to confirm or contradict an odd experience). AI systems, meanwhile, benefit from retrieval-augmented methods or specialized modules that can verify or refine uncertain outputs. In both cases, the capacity to anchor predictions in external references emerges as a principal shield against hallucinations, whether in clinical settings (where illusions may impair daily function) or in AI-driven applications (where fabricated information can cause tangible harm).

Finally, this comparative perspective highlights opportunities for ongoing interdisciplinary collaboration. Neuroscience and clinical psychology offer sophisticated models of predictive coding, showing how disturbances in error weighting lead to hallucinations; AI research contributes systematic approaches for grounding knowledge, checking consistency, and calibrating uncertainty. Each perspective holds lessons for mitigating hallucinations in the other. By framing hallucinations as a shared challenge in ``intelligent'' systems---biological or computational---our results pave the way for future endeavors that exploit these parallels to develop robust interventions and design principles, aiming to reduce the frequency and impact of ``confidently wrong'' perceptions across both fields.

\section{Limitations and Future Work}
\label{sec:limitations}

Although this comparative study highlights meaningful overlaps in the mechanisms of hallucinations across humans and AI, several caveats should be noted. First, the analysis relies on a broad, yet inherently selective, corpus of clinical neuroscience and machine-learning literature. Not every subtype of hallucination (for instance, purely olfactory phenomena in humans or code-level fabrications in AI) received the same depth of coverage. Second, attempts to parallel subjective experiences in humans with purely probabilistic outputs in AI necessarily bypass deeper questions about consciousness and emotional salience, which remain outside the scope of this paper \cite{fletcher2009perceiving, corlett2016prediction}.

A further limitation lies in the pace of both clinical and technological change. New insights into hallucinations—such as emerging psychotherapeutic modalities or neurobiological markers—may refine or contradict portions of our taxonomy. Likewise, large language models are evolving rapidly, suggesting that future architectures might introduce novel ways in which AI “hallucinates” \cite{lin2022truthfulqa, bommasani2021opportunities}. Finally, our emphasis on retrieval augmentation and metacognitive checks in AI, while instructive, is only one approach among many, and may not encompass all possible strategies for reducing confabulations in next-generation systems.

Looking ahead, bridging these limitations calls for more rigorous, interdisciplinary research. Studies could develop shared benchmarks that expose human participants and AI models to similarly ambiguous prompts, facilitating direct comparisons of error patterns. Incorporating embodied or multimodal AI systems—capable of integrating sensory feedback rather than relying solely on textual correlations—might also yield new insights into how grounding mitigates hallucinations. In the clinical realm, computationally inspired approaches to therapeutic intervention could refine techniques for recalibrating top-down priors in individuals prone to hallucinations. By examining these convergences and divergences more systematically, future work can deepen our understanding of predictive processing, reinforce more robust human–machine interaction, and ultimately foster better outcomes for those affected by erroneous perceptions or fabricated outputs.

\section{Conclusion}
\label{sec:conclusion}

This paper began by asking three central questions regarding hallucinations in humans and AI: how they emerge, whether AI hallucinations are merely imitations of human errors, and how insights from each domain can mitigate false perceptions. Our comparative analysis shows that both human brains and large language models rely on generative, predictive processes that fill in gaps with “best guesses.” In humans, this tendency arises when top-down priors overshadow weak or ambiguous sensory data, producing vivid but erroneous perceptions \cite{fletcher2009perceiving, corlett2016prediction}. In AI, the autoregressive paradigm can replace factual grounding with the most probable next token, creating confident yet unfounded statements \cite{lin2022truthfulqa, bommasani2021opportunities}. Far from a mere coincidence, these parallels underscore a fundamental predictive character in both domains.

From a clinical standpoint, hallucinations in humans often reflect a breakdown in balancing priors and sensory evidence—whether through cognitive-behavioral misalignments or neurochemical dysregulations \cite{friston2010freeenergy}. Interventions aimed at recalibrating these internal models, such as therapeutic feedback loops or targeted medication, can alleviate the severity of hallucinations. By analogy, AI systems benefit from retrieval augmentation, self-consistency checks, and cross-modal validation, which collectively ground text generation in verifiable data. 

Ultimately, these findings highlight that intelligence, whether biological or computational, is inherently probabilistic and generative. Hallucinations—neural or digital—arise when the predictive engine lacks robust corrective signals. By refining these correction mechanisms in both humans and AI, we can minimize the damage caused by confidently asserted errors, enhancing individual well-being, societal trust, and our broader understanding of intelligent behavior as a delicate balance between prior expectations and real-world evidence.

\end{document}